\documentclass[aps,prc,reprint,showpacs,groupedaddress,onecolumn]{revtex4}
\usepackage{bm}
\usepackage{amsmath}
\usepackage{amsfonts}
\usepackage{amscd} 
\usepackage{graphicx}
\usepackage{color}
\def\beq{\begin{equation}}
\def\eeq{\end{equation}}
\def\xf{\tilde{\mathbf{x}}}
\def\yf{\tilde{\mathbf{y}}}

\def\pf{\tilde{\mathbf{p}}}
\def\kf{\tilde{\mathbf{k}}}
\def\qf{\tilde{\mathbf{q}}}
\def\ff{\tilde{{f}}}
\def\gf{\tilde{{g}}}

\begin{document}

\title{The Light-Front Vacuum}

\author{Marc Herrmann}
\affiliation{Department of Physics and Astronomy, The University of
Iowa, Iowa City, IA 52242, USA}

\author{W.~N.~Polyzou}
\affiliation{Department of Physics and Astronomy, The University of
Iowa, Iowa City, IA 52242, USA}

\date{\today}

\begin{abstract}

{\parindent=0pt
{\bf Background:} The vacuum in the light-front representation of 
quantum field theory is trivial while vacuum in the equivalent canonical 
representation of the same theory is non-trivial.

{\bf Purpose:} Understand the relation between the vacuum in
light-front and canonical representations of quantum field theory
and the role of zero-modes in this relation.

{\bf Method:} Vacuua are defined as linear functionals on an algebra of 
field operators.  The role of the algebra in the definition of the vacuum
is exploited to understand this relation.

{\bf Results:} The vacuum functional can be extended from the
light-front Fock algebra to an algebra of local observables.  The
extension to the algebra of local observables is responsible for the
inequivalence.  The extension defines a unitary mapping between the
physical representation of the local algebra and a sub-algebra of the
light-front Fock algebra.
 
{\bf Conclusion:} There is a unitary mapping from the physical
representation of the algebra of local observables to a sub-algebra of
the light-front Fock algebra with the free light-front Fock vacuum.
The dynamics appears in the mapping and the structure of the sub-algebra.
This correspondence provides a formulation of locality and Poincar\'e
invariance on the light-front Fock space.
}

\end{abstract}

\maketitle

\section{Introduction}

The light-front \cite{Dirac:1949cp} representation of quantum field
theory has a number of
properties that make it advantageous for some applications,
particularly for applications involving electroweak probes of the
strong interaction and non-perturbative treatments of the strong
interaction \cite{Brodsky:1998}.  The properties of the light-front
representation of quantum field theory that lead to these advantages
are (1) there is a seven parameter subgroup of the Poincar\'e group
that is free of interactions (2) there is a three-parameter subgroup
of Lorentz boosts that is also free of interactions (3) the generator
of translations in the $x^-$ direction tangent to the light front is
free of interaction and satisfies a spectral condition (4) and the
vacuum of the interacting theory is the same as the vacuum of the free
field theory.

This is in contrast to the canonical formulation of quantum field
theory, where the six-parameter Euclidean group is free of
interactions, boosts depend on interactions and do not form a
subgroup, the spectrum of the translation generators is the real line,
and the vacuum of the interacting theory is not a vector in the free
field Fock space.

These differences have motivated many investigations into the 
nature of the light front vacuum 
\cite{Leutwyler:1970wn}
\cite{Schlieder:1972qr}
\cite{Chang:1972xt}
\cite{Tsujimaru:1997jt}
\cite{Nakanishi:1977}
\cite{Lenz:2000}
\cite{Werner:2006}
\cite{Beane:2013oia}.

While the light-front formulation of quantum field theory has
advantages in these applications, the predictions should be
independent of the representation of the theory.  It is found that
some light-front calculations require additional ``zero-mode''
contributions in order to maintain the equivalence with covariant
perturbation theory \cite{Bylev:1996} \cite{Bakker:2011zza}.

This work has two goals.  The first is to understand how the
vacuum can be trivial in one representation of field theory and not in
another equivalent representation of the same theory.  A second goal
is to understand the role of zero modes in the light-front vacuum.

It is instructive to consider what happens in the case of a free
scalar field theory.  The ground state of a harmonic oscillator is
uniquely determined by the condition that it is annihilated by the
annihilation operator.  Free scalar fields behave like infinite
collections of harmonic oscillators.  The analogy suggests that the
vacuum of the free-field theory is defined by the condition that it is
annihilated by a collection of annihilation operators, where the
annihilation operators are labeled by the three momentum.  Free fields
can be expressed as integrals over the 3-momentum that are linear in
the creation and annihilation operators.  A change of variables gives
an equivalent expression for the same field as an integral over the
light-front components of the momenta.  The resulting light-front and
canonical creation and annihilation operators are related by a
multiplicative factor, which is the square root of the Jacobian of the
variable change from the three momenta to the three light-front
components of the four momentum.  The multiplication of the
annihilation operator by a Jacobian should not impact the linear
equation that defines the vacuum, except possibly for momenta where
the Jacobian becomes singular.  This perspective suggests
that both representations of the free field theory should have the
same vacuum.

If two free scalar fields with different masses are restricted to the
light front, the masses do not appear in representations of fields, or
in the commutators of the creation and annihilation operators.  In
addition, the fields restricted to the light front are irreducible in
the sense that it is possible to extract both the creation an
annihilation operators directly from the field restricted to the light
front.  It follows from these properties that the two scalar fields
with different masses restricted to the light front and their
associated vacuum vectors are unitarily equivalent
\cite{Leutwyler:1970wn}. 

On the other hand, two free scalar fields with different masses are
related by a canonical transformation that expresses the annihilation
operator for the mass-1 field theory as a linear combination of a
creation and annihilation operators for the mass-2 field theory.  It
was explicitly shown by Haag \cite{Haag:1955ev} that, while this
canonical transformation can be represented by a unitary
transformation for finite numbers of degrees of freedom, this is no
longer true for canonical free fields.  For free fields of different
mass, the transformation is unitary if there is a momentum cutoff,
however when the cutoff is removed the generator of this unitary
transformation becomes ill-defined in the sense that it maps every
vector in the mass 1 representation of the Hilbert space, including
the vacuum, to a vector with infinite norm.  In this case the theories
are not related by a unitary transformation and the vacuua live in
different Hilbert space representations.  This is the well-known
problem of inequivalent representations of the canonical field algebra
for theories of an infinite number of degrees of freedom
\cite{segal:59}\cite{JvNeumann}\cite{Friedrichs:1953}.

These observations imply that the conventional mass-1
annihilation operator is related to the corresponding light-front
annihilation operator by a trivial Jacobian, which is unitarily
equivalent to the mass-2 light front annihilation operator.  This
operator is in turn related to the conventional mass-2 annihilation
operator by another Jacobian.  This suggests that the vacuua for both
theories are the same, or at least unitarily related, however this
contradicts the observation that the annihilation operators of the
conventional free field theories are related by a canonical
transformation that cannot be realized unitarily.

The virtue of free fields is that they are well understood.  Others
\cite{Bufalo:2014}\cite{Schlieder:1972qr}\cite{Werner:2006} have used
free fields to develop insight into various aspects of this problem.
Algebraic methods provide a resolution of the apparent inconsistency
discussed above. Algebraic methods were used in the seminal work of
Leutwyler, Klauder and Streit \cite{Leutwyler:1970wn} and Schlieder
and Seiler \cite{Schlieder:1972qr}.  This will be discussed in more
detail in the subsequent sections.

The problem is that the requirement that the vacuum is the state
annihilated by an annihilation operator does not give a complete
characterization of the vacuum of a local field theory.  In a local
field theory the vacuum is also a positive invariant linear functional
on the field algebra.  The relevant algebra is the algebra of local
observables, which is not the same as the canonical equal-time algebra
or the algebra generated by fields restricted to a light front.  While
the desired positive linear functional can be expressed by taking
vacuum expectation values of elements of the algebra with the vacuum
defined by the annihilation operator, the definition of the vacuum
functional also depends on the choice of algebra.  Specifically,
functionals that agree on a sub-algebra do not have to agree on the
parent algebra.  The physically relevant algebra for a quantum field
theory must be large enough to be Poincar\'e invariant and to contain
local observables.  Both the canonical and light-front algebras (which
are defined later) are irreducible in the sense that they can be used to
build both the Hilbert space and operators on the Hilbert
space, but they are not closed under Poincar\'e transformations and do
not contain local observables.  While they are not sub-algebras of the
local algebra, for the case of free fields the irreducibility allows
the linear functional that defines the vacuum on these algebras to be
extended to the local algebra.  In the case of the canonical
equal-time algebra the extension is essentially unique
\cite{PhysRev.117.1137}\cite{Araki:1971yj}, while in the case of the
light-front algebra there are many extensions that lead to
inequivalent representations of the local algebra.  These extensions
define a unitary map that relates local algebra and physical vacuum
to a sub-algebra of the light front algebra with the free Fock vacuum.

In the light-front case the extension to the local algebra requires
additional attention to what happens when the $+$ component of the
momentum is 0.  For free fields the extension to the local algebra
regularizes apparent singularities that are associated with $p^+=0$.  

The algebraic methods discussed for free fields can also be applied to
interacting fields by first representing them using a Haag expansion
\cite{Glaser:1957}\cite{Greenberg:1965}\cite{Haag:1955ev} as a series
in a complete set of normal products of asymptotic ({\it in} or {\it
out}) fields.  The asymptotic fields are free fields, and each of
them can be expressed as an extension of a free field restricted to the
light front.  This results in an extension of the light-front algebra
to the local algebra generated by the interacting field.

The coefficients of the Haag expansion of the interacting Heisenberg
field are invariant (covariant) functions
\cite{Glaser:1957}. Additional properties of these functions follow
because the Heisenberg fields and asymptotic fields are
operator-valued tempered distributions.  When the asymptotic fields in
the Haag expansion are replaced by the extensions of free fields on
the light-front to the asymptotic fields, the result is an expansion
of the Heisenberg fields in terms of normal products of free fields
restricted to a light front.  The coefficient functions in this
expansion regulate the $p^+=0$ behavior of the free light-front
fields.

More singular behavior can occur in operators, like Poincar\'e
generators, that involve products of fields at the same point on the
light front.  These products are ill defined and a renormalization is
necessary for them to make sense as operators.  Divergences appear for
both large momenta as well as $p^+=0$. The resulting finite theory
needs to be independent of the orientation of the light front.
Invariance under change of orientation of the light front is
equivalent to rotational covariance of the theory
\cite{Karmanov:1976}\cite{Karmanov:1988}
\cite{Fuda:1990}\cite{Fuda:1991nn}
\cite{Fuda:1994uv}\cite{Polyzou:1999}. Since $p^+=0$ for one light
front correspond to some component of $p$ becoming infinite with a
different light front,  rotational covariance necessarily puts
important constraints on the renormalization. 

This also suggests that the zero mode issue is more
complicated in 3+1 dimensional theories than 1+1 dimensional theories,
where rotational covariance plays no role.  Specifically, the
extension to the local algebra must recover both the rotational
covariance and locality of the theory.

In the next section we define our notation and conventions.  In
section three we discuss inequivalent representations. In section four
we discuss the-light front vacuum.  In section five we introduce four
different field algebras that we use in this paper.  Section six
discusses the light-front Fock algebra.  In section seven we discuss
the meaning of equivalent theories.  In section eight we discuss
extensions of the light-front Fock algebra.  In section 9 we discuss
dynamical theories.  Zero modes are discussed in section ten.  The
results are summarized in section eleven.
 
\section{Notation}

This section defines the notation and conventions that will be used in the
remainder of the paper.  The signature of the Minkowski metric is
$(-,+,+,+)$.  Space-time components of four vectors, $x$, have Greek
indices
\beq
x^{\mu}  := (x^0,\mathbf{x}). 
\label{n.1}
\eeq
A light front is a three-dimensional hyper-plane in Minkowski space
satisfying 
\beq
x^+:= x^0 + \hat{\mathbf{n}}\cdot{\mathbf{x}}= 0
\label{n.2}
\eeq
where $\hat{\mathbf{n}}$ is a fixed unit vector.  Points on
the light front have either a space-like or light-like separation.
Coordinates of points on the light front are
\beq
\tilde{\mathbf{x}} = (x^-,\mathbf{x}_{\perp})
\label{n.3}
\eeq
where 
\beq
x^{-}= x^0 - \hat{\mathbf{n}}\cdot{\mathbf{x}}
\qquad
\mbox{and}
\qquad
\mathbf{x}_{\perp} = \mathbf{x} - \hat{\mathbf{n}}
(\hat{\mathbf{n}}\cdot{\mathbf{x}}).
\label{n.4}
\eeq
The light-front components of a four vector $x$ are
\beq
x= (x^+ , x^- ,\mathbf{x}_{\perp}).
\label{n.5}
\eeq
The Lorentz invariant scalar product of two four vectors 
in terms of their light-front components is
\beq
x \cdot y = - {1 \over 2} x^+y^- - {1 \over 2}x^- y^+ + 
\mathbf{x}_{\perp}\cdot \mathbf{y}_{\perp}.
\label{n.6}
\eeq

The Fourier transform of a Schwartz test function $f(x)\in S(\mathbb{R}^4)$ 
is 
\beq
{f}(p) 
= {1 \over (2 \pi)^{2}} \int 
e^{- i p \cdot x } f(x) {1 \over 2} dx^+ dx^- d^2 x_{\perp} 
=
{1 \over (2 \pi)^{2}} \int 
e^{- i p \cdot x } f(x) d^4 x
\label{n.7}
\eeq
\beq
{f}(x) = 
{1 \over (2 \pi)^{2}} \int 
e^{ i x \cdot p } {f}(p) {1 \over 2} dp^+ dp^- d^2 p_{\perp}
=
{1 \over (2 \pi)^{2}} \int 
e^{ i x \cdot p } {f}(p) d^4 p
\label{n.8}
\eeq
where $d^4x = {1 \over 2} dx^+ dx^- d^2 x_{\perp}$.   A 
``mathematically imprecise'' notation is used to represent
functions $f(x)$ and their Fourier transforms $f(p)$,
which are related by (\ref{n.7}) and (\ref{n.8}).  In what
follows $p,k$ and $q$ represent momentum variables and $x,y$ and $z$ represent
coordinate variables.   It is useful to define
\beq
\pf := (p^+ ,\mathbf{p}_{\perp})
\qquad
\xf \cdot \pf := -{1 \over 2} x^-p^+ + 
\mathbf{x}_{\perp}\cdot \mathbf{p}_{\perp} 
\label{n.9}
\eeq
and 
\beq
d\xf := dx^- d\mathbf{x}_{\perp}
\qquad
d\pf := dp^+ d\mathbf{p}_{\perp}.
\label{n.10}
\eeq
In this notation the 
three-dimensional Fourier transform of functions
$\ff (\xf)$ in the light-front 
variables are 
\beq
\ff(\pf) = 
{1 \over 2^{1/2}(2 \pi)^{3/2}} \int 
e^{- i\pf \cdot \xf } 
\ff(\xf ) 
d\xf
\label{n.11}
\eeq
and
\beq
\ff (\xf ) = 
{1 \over 2^{1/2} (2 \pi)^{3/2}} \int 
e^{ i\pf \cdot \xf } 
\ff (\pf) 
d\pf.
\label{n.12}
\eeq
The $\tilde{}$ indicates functions supported on the light
front and their Fourier transforms.

The light front is invariant under a seven-parameter subgroup, called
the light-front kinematic subgroup of the Poincar\'e group.  This
subgroup is generated by the three-parameter subgroup of translations
tangent to the light front,
\beq
\tilde{\mathbf{x}}: \to \tilde{\mathbf{x}}'=
\tilde{\mathbf{x}} +\tilde{\mathbf{a}}
\label{n.13}
\eeq
a three-parameter subgroup of light-front preserving boosts,  
\beq
x^+ \to  x^{+\prime} = q^+ x^+
\qquad
{\mathbf{x}}_{\perp} \to {\mathbf{x}}'_{\perp}=  
{\mathbf{x}}_{\perp} + {\mathbf{q}}_{\perp}x^+
\label{n.14}
\eeq
\beq
x^- \to x^{-\prime}= {1 \over q^+}
(x^- + {\mathbf{q}}_{\perp}^2 x^+ + 2 {\mathbf{q}}_{\perp}\cdot
{\mathbf{x}}_{\perp})
\label{n.15}
\eeq
and rotations about the $\hat{\mathbf{n}}$ axis.  

Light-front boosts applied to points on the light front
only transform $x^-$:
\beq
{\mathbf{x}}_{\perp} \to {\mathbf{x}}'_{\perp}=  
{\mathbf{x}}_{\perp}  
\label{n.16}
\eeq
\beq
x^{-\prime}= {1 \over q^+}
(x^- + 2 {\mathbf{q}}_{\perp}\cdot
{\mathbf{x}}_{\perp}),
\label{n.17}
\eeq
however the conjugate momentum variables $\tilde{\mathbf{p}}$
can take on any value under these transformations provided
$p^+\not=0$:
\beq
p^+ \to  p^{+\prime} = q^+ p^+
\qquad
{\mathbf{p}}_{\perp} \to {\mathbf{p}}'_{\perp}=  
{\mathbf{p}}_{\perp} + {\mathbf{q}}_{\perp}p^+ .
\label{n.18}
\eeq
When $p^+=0$ the light-front boosts leave $\tilde{\mathbf{p}}$ 
unchanged. 

The light-front inner product is defined by
\beq
(\tilde{f},\tilde{g}) = 
\int {d\pf \theta(p^+) \over p^+} \tilde{f}^*(\pf)
\tilde{g}(\pf ).
\label{n.19}
\eeq 
This inner product is invariant with respect to the kinematic subgroup
because (1) the measure is invariant and (2) $\pf \to \pf'$ does to
involve $p^-$. This inner product has a logarithmic singularity
for functions $\tilde{f}(\pf )$ that are non-zero at $p^+=0$.

\section{Inequivalent representations}

Free fields look like collections of uncoupled harmonic oscillators.
For a single oscillator the Hamiltonian in dimensionless 
variables is
\beq
H = {1 \over 2} (x^2 + p^2) = a^{\dagger} a +{1 \over 2}
\label{irep:1}
\eeq
where
\beq
[x,p]=i  \qquad a := (x+ip)/\sqrt{2} \qquad [a,a^{\dagger}]=1.
\label{irep:2}
\eeq
The equation 
\beq
a \vert 0 \rangle =0.
\label{irep:3}
\eeq 
determines the ground state $\vert 0 \rangle$ 
of the oscillator.

The canonical transformation 
\beq
x \to x' = \alpha  x; p \to p'= {p \over \alpha } 
\label{irep:4}
\eeq
preserves $[x,p]=[x',p']=i$, and leads to a 
linear relation between the original and transformed 
annihilation operators: 
\beq
a' ={1 \over 2} ({\alpha}+{1 \over \alpha} ) a + 
{1 \over 2} (\alpha -{1 \over \alpha}) a^{\dagger}
= \cosh (\eta) a + \sinh (\eta) a^{\dagger}
.
\label{irep:5}
\eeq
This canonical transformation can be implemented by 
the unitary operator $U$:
\beq
a' = U a U^{\dagger}
\qquad 
U = e^{{\eta \over 2 }(aa-a^{\dagger}a^{\dagger})} =  e^{iG} .
\label{irep:6}
\eeq
The transformed Hamiltonian 
\beq
H' = {1 \over 2} (x^{\prime 2} + p^{\prime 2}) = a^{\prime \dagger} a' +{1 \over 2}
\label{irep:7}
\eeq
has the same eigenvalues as $H$. The transformed ground state
vector is related to the original ground state vector by
\beq
\vert 0' \rangle = U \vert 0 \rangle . 
\label{irep:8}
\eeq

The canonical transformation (\ref{irep:4}) is the single-oscillator 
version of the canonical transformation that changes the mass 
in a free field theory.

The canonically conjugate operators in a free scalar field theory 
have the form
\beq
\phi (x) = (2 \pi)^{-3/2} \int d\mathbf{p} 
{1 \over \sqrt{2 \omega_m (\mathbf{p})}} (a(\mathbf{p}) e^{i p \cdot x} 
+ a^{\dagger}(\mathbf{p}) e^{-i p \cdot x}) 
\label{irep:9}
\eeq
\beq
\pi (x) = -i (2 \pi)^{-3/2} \int d\mathbf{p} 
 \sqrt{{ \omega_m (\mathbf{p}) \over 2}} (a(\mathbf{p}) e^{i p \cdot x} 
- a^{\dagger}(\mathbf{p}) e^{-i p \cdot x})
\label{irep:10}
\eeq
where $x$ is restricted to $t=0$ and 
$p^0= \omega_m (\mathbf{p}) = \sqrt{\mathbf{p}^2 +m^2}$ is the energy.

The canonical transformation that changes the mass $m \to m'$ involves
multiplying the integrand of $\phi (x)$ by
$\alpha (\mathbf{p}) = \sqrt{{\omega_{m} (\mathbf{p}) \over\omega_{m'}
(\mathbf{p})}} $
and the integrand of $\pi (x)$ by $1/\alpha (\mathbf{p})$.  This has
the same structure as the single oscillator canonical transformation
(\ref{irep:4})
except in this case the parameter $\alpha$ depends on the momentum.

As in the case of the single oscillator, define
$\eta (\mathbf{p})$ by 
\beq
\cosh (\eta (\mathbf{p})) = {1 \over 2} (\alpha (\mathbf{p}) + 
{1 \over \alpha ( \mathbf{p})})
\qquad  
\sinh (\eta(\mathbf{p})) = {1 \over 2} (\alpha (\mathbf{p}) - 
{1 \over \alpha ( \mathbf{p})}) .
\label{irep:11}
\eeq
By analogy with
(\ref{irep:6}), the formal generator $G$ of the canonical transform 
that changes the mass is 
\beq 
G \to -i \int {\eta (\mathbf{p}) \over 2} \left (a(\mathbf{p}) a(\mathbf{p}) -
a^{\dagger}(\mathbf{p}) a^{\dagger}(\mathbf{p})\right ) d \mathbf{p}.
\label{irep:12}
\eeq
A simple calculation shows
\beq
\Vert G \vert 0 \rangle \Vert^2 = {1 \over 4} \int \eta (\mathbf{p})^2
d\mathbf{p} \delta (0) = \infty
\label{irep:13}
\eeq
which implies that the domain of $G$ is empty.

The consequence is that the canonical transformation relating free
field theories with different masses cannot be realized by a unitary
transformation.  The second is that the canonical vacuum vectors
characterized by
\beq
a_i (\mathbf{p}) \vert 0 \rangle_i =0
\qquad i \in \{ 1,2 \}
\label{irep:14}
\eeq   
are not unitarily related.  The existence of unitarily inequivalent
representations of the canonical commutation for systems of an
infinite number of degrees of freedom is well known \cite{segal:59}.
The example above was discussed by Haag
\cite{Haag:1955ev}.

The interest in this example is that, in contrast to canonical
equal-time fields,  for free fields restricted to
a light front the vacuum vectors and annihilation operators
for free fields of different mass are unitarily related
\cite{Leutwyler:1970wn}.

\section{Light-front vacuum}

In light-front field theory the generator $P^+$ of translations in the
$x^-$ direction is a kinematic operator satisfying the spectral
condition, $P^+\geq 0$.  The dynamical operator is $P^-$, which
generates translations normal to the light front, is the
light-front Hamiltonian.  It can formally be expressed as the sum of a
non-interacting term and an interaction
\beq
P^-=P^-_0 + V .
\label{lfv:1}
\eeq
Kinematic translational invariance of $P^-$ and $P^-_0$
on the light-front requires
that the interaction commutes with $P^+$:
\beq
[P^+,V]=0 .
\label{lfv:2}
\eeq
It follows that
\beq
P^+ V \vert 0 \rangle = V P^+ \vert 0 \rangle =0  
\label{lfv:3}
\eeq
for a light-front translationally invariant  
vacuum.

Because $P^+$ is kinematic and satisfies a spectral condition $(P^+
\geq 0)$, it follows that $P^+ = \sum_i P_i^+ \geq 0$.  
If $V$ can be expressed as a kernel integrated against creation
and annihilation operators, then the coefficient of the pure creation
terms in the interaction must vanish unless $p_i^+=0$ for each
creation operator.  If this kernel is a continuous function of $p^+$
the interaction will necessarily leave the vacuum is unchanged.
This does not rule out singular contributions at $p^+=0$ which
are associated with zero-modes of the theory.

For the special case of a free field of mass $m$ the light-front
representation of the field is obtained by changing the integration
variable from $\mathbf{p}$ to the three light-front components of the
four momentum.  The result is
\beq
\phi (x) = (2 \pi)^{-3/2} \int 
{dp^+ d\mathbf{p}_{\perp}\theta (p^+) 
\over \sqrt{2 p^+} }
({{\sqrt{\omega_m (\mathbf{p})} \over p^+}} a(\mathbf{p}) e^{i p \cdot x} + 
{\sqrt{\omega_m (\mathbf{p}) \over p^+}}
a^{\dagger}(\mathbf{p}) e^{-i p \cdot x}).
\label{lfv:4}
\eeq
This leads to the following relation between the
light-front and canonical annihilation operators
\beq
a_{lf} (\tilde{\mathbf{p}}):= a_{lf}(p^+,\mathbf{p}_{\perp} ) = 
\sqrt{{\omega_m (\mathbf{p}) \over p^+}} a(\mathbf{p})
\label{lfv:5}
\eeq
which satisfy
\beq
[a_{lf}(p^+,\mathbf{p}_{\perp} ),  a^{\dagger}_{lf}(q^+,\mathbf{q}_{\perp} ) 
] = \delta (\tilde{\mathbf{p}} - \tilde{\mathbf{q}})
= \delta (p^+-q^{+})\delta (\mathbf{p}_{\perp} -\mathbf{q}_{\perp}) . 
\label{lfv:6}
\eeq
As in the canonical case, the light front-vacuum is characterized by
the condition that it is annihilated by the annihilation operator
\beq
a_{lf} ((p^+,\mathbf{p}_{\perp} )\vert 0 \rangle = 
{\sqrt{\omega_m (\mathbf{p}) \over p^+}} a(\mathbf{p}) \vert 0 \rangle 
=0
\label{lfv:7}
\eeq
If this field is restricted to the light front $x^+=0$ the
mass-dependent factors disappear both from the field (\ref{lfv:4}) and the
commutation relations (\ref{lfv:6}) of $a_{lf}$ and $a_{lf}^{\dagger}$.
In this case vacuum expectation values of a product of fields
smeared with functions supported on the light front are all
identical, independent of mass.  

The equation (\ref{lfv:7}) is the same for all free scalar fields.

\section{Field algebras}

An obvious question is how are these different characterizations of
the vacuum of a free field of mass $m$ are related.  If the vacuum is
uniquely characterized by the annihilation operator, like it is for a
single harmonic oscillator, the arguments of section 3 imply that the
vacuua for free fields with different masses are inequivalent.  On the
other hand, the arguments of the previous section imply that the
light-front vacuum is identical to the canonical vacuum and light
front vacuum vectors for different masses are unitarily related.  The
resolution of these apparently conflicting results can be understood
by giving up the assumption that the vacuum is uniquely characterized
by the annihilation operator.

The critical observation is that each of the vacuua discussed above
are implicitly defined on different algebras of operators.  The
relevant algebra for a local quantum field theory must be invariant
with respect to Poincar\'e transformation and contain observables
associated with arbitrarily small spacetime volumes.  These two
conditions are {\it not} satisfied by both the light-front and
canonical fixed-time field algebras, even though both of these
algebras are irreducible.

This suggests that the characterization of the vacuum by an annihilation
operator is incomplete.  The vacuum functionals generated by both the
light-front and canonical annihilation operators require non-trivial
extensions in order to define linear functionals on a Poincar\'e
invariant algebra of local observables. 

Different algebras are generated by smeared fields of the form
\beq
\phi (f) = \int \phi (x) f(x) d^4 x 
\label{al.1}
\eeq
where we limit our considerations to the case that 
$\phi (x)$ is a scalar field.  The field algebra consist of
polynomials in $\phi (f)$ or $e^{i \phi (f)}$.  The difference being
that the exponential form is bounded so there are no issues with
domains, however this distinction is not important in this paper.
What is relevant is that for a given scalar field there are different
algebras that are distinguished by different choices of the space of
test functions.

It is useful to identify the following four free field algebras.

We call the algebra generated by local observables, i.e. fields smeared 
with Schwartz functions in four space-time variables,
\beq
\{ \phi (f) \vert f(x) \in S(\mathbb{R}^4) \},
\label{al.2}
\eeq
the local algebra.  

The algebra generated by fields smeared with Schwartz functions in
three coordinates on the light front,
\beq
\{ 
\phi (f) \vert f(x)= \delta (x^+) {f} (\tilde{\mbox{x}}),
\qquad f(\tilde{\mbox{x}}) \in S(\mathbb{R}^3)  
\}, 
\label{al.3}
\eeq
is called the light-front algebra.  The algebra generated by  
fields and their time derivatives smeared with Schwartz functions in 
three spatial coordinates,
\beq
\{ \phi (f) \vert f(x)= \delta (t) f (\mathbf{x}),
-\dot{\delta}(t) f (\mathbf{x})
\qquad f (\mathbf{x}) \in S(\mathbb{R}^3)  \}, 
\label{al.4}
\eeq
is called the canonical algebra.  Integrating the field over these
test functions gives the canonically conjugate $\phi (\cdot)$ and
$\pi (\cdot)$ fields restricted to the $t=0$ hyper-plane.  The algebra
generated by fields smeared with Schwartz functions in three
light-front coordinates, restricted to have zero $x^-$ integral
\beq
\{ \phi (f) \vert f(x)= \delta (x^+) {f} (\tilde{\mathbf{x}})
\qquad f(\tilde{\mathbf{x}}) \in S(\mathbb{R}^3); 
\int f(\tilde{\mathbf{x}}) dx^- =0  \}, 
\label{al.5}
\eeq
is called the Schlieder-Seiler algebra\cite{Schlieder:1972qr}.  The
restriction on the test functions means the their Fourier transform
vanishes at $p^+=0$, which makes the light-front inner product
(\ref{n.19}) finite.  

The local algebra is the only one of these four algebras that is
preserved under the Poincar\'e group and contains observables that can
be localized in any space-time volume.  The light-front and
Schlieder-Seiler algebras are preserved under the kinematic subgroup
of the light-front and the canonical algebra is preserved under the
three-dimensional Euclidean group.  The reason for distinguishing the
light-front and Schlieder-Seiler algebras is that the light-front Fock
vacuum is not defined on the light-front algebra while it is
well-behaved on the Schlieder-Seiler algebra.

A vacuum on any of these algebras is a positive linear functional $L$.
The positivity condition means that
\beq
L[A^{\dagger} A] \geq 0
\label{al.6}
\eeq
for any element $A$ of the algebra.  Each such positive linear
functional $L[\cdot]$, can be used to construct a Hilbert-space
representation of the algebra with inner product
\beq
\langle B \vert A \rangle =  L[B^{\dagger} A] .  
\label{al.7}
\eeq
Formally, the Hilbert space representation is constructed by
identifying vectors whose difference has zero norm and completing the
space by adding Cauchy sequences.  This is the standard GNS
construction which is discussed in many texts\cite{araki-bk} .  There
are additional constraints on the linear functional for it to
represent a vacuum vector.  One of these properties is that when
$L[\cdot]$ is invariant with respect to the group that preserves the
algebra, then the Hilbert space representation of the group is
unitary.

Another property of all of these algebras is that they are
irreducible.  This means that any bounded linear operators on the Hilbert
space can also be formally expressed in terms of operators in the
algebra.

Acceptable candidates for vacuum vectors are invariant positive linear
functionals on the algebra.  While the vacuum expectation value
of an element of the algebra with the vector 
annihilated by the annihilation operator defines a linear functional,
the definition of the functional also depends on the
algebra.  Specifically, since positivity on a sub-algebra does not
imply positivity on the parent algebra, and a restricted symmetry on a
sub-algebra does not imply the full symmetry on the parent algebra, it
follows that the characterization of the vacuum as a positive
invariant linear functional depends on the choice of algebra.  While
this is different than the characterization in terms of annihilation
operators, it does not preclude the possibility of using a vacuum
characterized by a particular annihilation operator, but the linear
functional must be extended to the local algebra.

While the canonical, light-front, and Schlieder-Seiler algebras for
free fields are not sub-algebras of the local free-field algebra,
these algebras are related to the local algebra by the standard
expressions for the field operator, $\phi(x)$.

While local algebras exist for free or interacting Heisenberg fields,
because fields are operator-valued distributions the other three
algebras may not exist in general because the ``test functions''
(\ref{al.3},\ref{al.4},\ref{al.5}) have a distributional component.
However for the case of free fields these restrictions are defined.
For the case of free fields, the formal representation of the fields
in terms of creation and annihilation operators provides explicit
relations among these four algebras.

\section{Light-front Fock algebra}

In this section we give a more complete description of the light-front
and Schlieder-Seiler algebras and review some properties
\cite{Leutwyler:1970wn} \cite{Schlieder:1972qr} of these algebras.

The light-front Fock algebra is the algebra of free field operators
smeared with Schwartz functions in the light-front coordinates $\xf$
with $x^+=0$.  A free scalar field of mass $m$ expressed in terms of
light-front coordinates has the form:
\beq 
\phi (x) = {1 \over (2 \pi)^{3/2}} \int {d\pf \theta
  (p^+) \over \sqrt{2p^+}} (a_{lf} (\tilde{\mathbf{p}}) e^{i p \cdot x}+
a_{lf}^{\dagger} (\tilde{\mathbf{p}}) e^{- i p \cdot x})
\label{f.1}
\eeq
where the mass $m$ only enters in $e^{\pm i p \cdot x}$ 
for $x^+\not=0$:
\beq 
p^- = {\mathbf{p}_{\perp}^2 + m^2 \over p^+} .
\label{f.2}
\eeq
On the light front, when $x^+=0$, all information about the mass is 
lost.   The creation and annihilation operators satisfy
\beq
[a_{lf} (\tilde{\mathbf{p}}),a_{lf}^{\dagger} (\tilde{\mathbf{k}})]
=\delta (\tilde{\mathbf{p}} - \tilde{\mathbf{k}}) ,
\label{f.3}
\eeq
which is also independent of mass.

Equation (\ref{f.2}) is where the free field dynamics enters; 
it extends the field restricted to a light front to a field,
$\phi(x)$, on the local algebra that satisfy the Klein-Gordon equation
\cite{Jerzy}
\beq
(\Box - m^2) \phi (x)=0.
\label{f.4}
\eeq
Each mass defines a distinct and inequivalent extension of the
light-front free-field algebra to different local algebras.

The Fourier transform of the field restricted to the light front can be 
decomposed into terms with positive and negative values of $\mathbf{p}^+$:
\[
\tilde{\phi}(\pf) :=
{1 \over (2)^{1/2}(2 \pi)^{3/2}} \int d\xf 
e^{-i \pf \cdot \xf}  
{\phi}(0,\xf) =
\]
\beq 
\theta (p^+) \tilde{\phi}(\pf)+ \theta (-p^+)  \tilde{\phi}(\pf)=
\theta (p^+) 
 \sqrt{{1 \over p^+}}a_{lf}(\pf) +
\theta (-p^+) 
\sqrt{{- 1\over  p^+}}a_{lf}^{\dagger}(-\pf) .
\label{f.5}
\eeq
This decomposition only makes sense for $p^+ \not= 0$. This means 
that this decomposition is only defined on the Schlieder-Seiler algebra, 
where the test functions vanish at $p^+=0$.

This decomposition can be used to separate the creation 
and annihilation operators:
\beq
a_{lf}(\pf) =
\theta (p^+) 
\sqrt{{p^+ \over 2}}
{1 \over (2 \pi)^{3/2}} \int d\xf
e^{-i \pf \cdot \xf}  
{\phi}(0,\xf) 
\label{f.6}
\eeq
and 
\beq
a_{lf}^{\dagger}(\pf)=
\theta (p^+) 
\sqrt{{p^+ \over 2 }}
{1 \over (2 \pi)^{3/2}} \int d\xf 
e^{i \xf \cdot \pf}  
{\phi}(0,\xf) .
\label{f.7}
\eeq
This property, that one can extract both creation and annihilation
operators from the field restricted to the light front, is not shared
with fields restricted to a space-like hyper-plane.  On a space-like
hyper-plane, both the field and its time derivative (which requires
knowing about the field off of the fixed-time hyper-plane)
are needed to independently extract the
creation and annihilation operators.  

Equation (\ref{f.5}) implies that the field restricted to the light 
front can be decomposed into parts corresponding to the sign of $p^+$ 
in the Fourier transform
\beq
\tilde{\phi}(\xf ):= \phi (0,\xf) = \phi^+ (0,\xf) + \phi^- (0,\xf)
\label{f.8}
\eeq
where 
\beq
\tilde{\phi}^{\pm} (\xf) = {1 \over (2)^{1/2}(2 \pi)^{3/2}}\int d\pf e^{i \pf\cdot \xf}
\tilde{\phi}^{\pm} (\pf)
\label{f.9}
\eeq
and
\beq
\tilde{\phi}^+ (\pf):=
{\theta (p^+) 
\over 
\sqrt{p^+}}a_{lf}(\pf) 
\qquad
\tilde{\phi}^- (\pf):=
{\theta (-p^+) 
\over 
\sqrt{-p^+}}a_{lf}^{\dagger}(-\pf) .
\label{f.10} 
\eeq
Next we argue that the spectral condition on $p^+$ leads to an 
algebraic definition of normal ordering.  By this we mean that 
vacuum expectation values are not explicitly needed to define the 
normal product of light front field operators.

To see this note that translational covariance of the field 
\beq
e^{i P^+ a^-} \tilde{\phi} (\mathbf{\tilde{x}}) 
e^{- i P^+ a^-} = \tilde{\phi} (\mathbf{\tilde{x}} + a^-).
\eeq
implies
\beq
[P^+,\tilde{\phi}^\pm (\pf)] =
\mp \vert p^+\vert  \tilde{\phi}^\pm (\pf).
\label{f.12}
\eeq
It follows that if $\vert  q^+ \rangle$ is an eigenstate
of $P^+$ with eigenvalue $q^+$ then
\beq
P^+\tilde{\phi}^\pm (\pf) \vert q^+ \rangle =
([P^+ , \tilde{\phi}^\pm (\pf)] + \tilde{\phi}^\pm (\pf)P^+) 
\vert q^+ \rangle =
(q^+ \mp \vert p^+\vert ) \tilde{\phi}^\pm (\pf) \vert q^+ \rangle . 
\label{f.13}
\eeq
This shows that application of $\tilde{\phi}^+ (\pf)$ to any eigenstate of
$P^+$ results in an eigenstate of $P^+$ with a lower eigenvalue of
$P^+$.  Since kinematic translational invariance in the $x^-$
direction implies that any vacuum is an eigenstate of $P^+$ with
eigenvalue 0, the spectral condition $P^+\geq 0$ implies that
$\tilde{\phi}^+ (\pf)$ must annihilate that vacuum as long as the support of
$\pf$ does not contain the point $p^+ =0$.  This is always the case for the
Schlieder-Seiler algebra.  This is a consequence of the algebra - it
holds for any Hilbert space representation of the light-front
Schlieder-Seiler algebra.  It is important to note that the only
property of the vacuum that was used was translational invariance,
which is a property of any light-front invariant vacuum.  The
decomposition (\ref{f.8}) makes no use of the vacuum.
 
It follows that there is an algebraic notion of normal ordering on the
free field light-front Schlieder-Seiler algebra.  The rule is to
decompose every field operator $ \tilde{\phi} (\ff) = \tilde{\phi}^- (\ff ) +
\tilde{\phi}^+
(\ff)$, and then move all of the $\tilde{\phi}^+ (\ff ) $ parts of the field
to the right of the $\tilde{\phi}^- (\ff)$ parts of the fields.  We use the
standard double dot $:\, :$ notation to indicate {\it algebraic}
normal ordering.  We refer to it as ``algebraic'' because a vacuum is
not needed to make the decomposition (\ref{f.8}).

The light-front vacuum for free fields is uniquely
determined by the algebraic normal ordering on the Schlieder-Seiler algebra.
To see this note that it follows from the decomposition (\ref{f.10})
that
\beq
U(\ff) = e^{i \tilde{\phi}(\ff)} =
e^{i \tilde{\phi}_+(\ff)+ i \tilde{\phi}_- (\ff) } =
e^{i \tilde{\phi}_-(\ff)}e^{i\tilde{\phi}_+ (\ff) }
e^{{1 \over 2} (f,f)}=
:e^{i \tilde{\phi}(\ff)}:
e^{{1 \over 2} (\ff,\ff)}
\label{f.14}
\eeq
which expresses $U(\ff)$ as an algebraically 
``normal ordered operator'' 
multiplied by the known coefficient function, $e^{{1 \over 2} (\ff,\ff)}$.
If the vacuum expectation value 
of the normal product $:e^{i \phi(\ff)}:$ is $1$, then the vacuum functional 
on this algebra by is given by
\beq
\langle 0 \vert U(\ff) \vert 0 \rangle = 
e^{{1 \over 2} (\ff,\ff)}.
\label{f.15}
\eeq
Since $(\ff,\ff)$ is ill-defined for functions that do not vanish at
$p^+=0$, this vacuum is only defined on the Schlieder-Seiler algebra.
Furthermore, on the Schlieder-Seiler algebra the vacuum expectation
value of the normal product $:e^{i \tilde{\phi}(\ff)}:$ is 1 as a result
of (\ref{f.13}) and the fact that the Schlieder-Seiler test functions
vanish for $p^+=0$.

It follows that for free fields of any mass, the Schlieder-Seiler
algebras are unitarily equivalent \cite{Leutwyler:1970wn}.  
This is because the vacuum
expectation values of any number of smeared fields are identical.

The formal irreducibility of this algebra follows because it  
has the structure of a Weyl algebra \cite{Klauder:1969}\cite{Araki:1963},  
but
in this case the algebra has no local observables and the class
of test functions is too small to determine any dynamical information.
The Weyl structure is contained in the unitary operators
\beq
U(\ff)= e^{i \tilde{\phi} (\ff)}. 
\label{f.16}
\eeq
Products of field operators are replaced by 
products of bounded operators of the form (\ref{f.16}).
Products of two such operators can be computed using the 
Campbell-Baker-Hausdorff theorem\cite{Naimark:1982}.   The result is 
\beq
U(\ff ) U(\gf ) = U (\ff+\gf) e^{-{1 \over 2}[\tilde{\phi} (\ff),\tilde{\phi} (\gf)]} =
U (\ff+\gf) e^{-{1\over 2}
((\tilde{f},\tilde{g})-(\tilde{g},\tilde{f}))} ,
\label{f.17}
\eeq
which is another operator of the same form multiplied by 
the scalar coefficient, $e^{-{1\over 2}
((\tilde{f},\tilde{g})-(\tilde{g},\tilde{f}))} $. 

To  put (\ref{f.17}) in the form of a Weyl algebra 
first decompose the Fourier transform of a real Schlieder
Seiler test function into
real and imaginary parts;
$\ff = \ff_r + i \ff_i$.  Defining
\beq
U(\ff)= U(\ff_r,\ff_i)
\label{f.18}
\eeq
equation (\ref{f.17}) becomes
\beq
U(\tilde{f}_r,\tilde{f}_i) U(\tilde{g}_r,\tilde{g}_i) =
U(\tilde{f}_r+\tilde{g}_r,\tilde{f}_i+\tilde{g}_i) e^{-{i \over 2}
((\tilde{f}_r,\tilde{g}_i)-(\tilde{g}_r,\tilde{f}_i))}.
\label{f.19}
\eeq
This has the same form as the Weyl algebra for the canonical
fields if we use
\beq
U(f,g) := e^{i \phi (f) + i \pi (g)} 
\label{f.20}
\eeq
where the light front-inner product is replaced by the
ordinary $L^2(\mathbb{R}^3)$ inner product. 

While the light-front inner product (\ref{n.19}) is singular for
functions that do not vanish at $p^+=0$, the difference
$(\tilde{f}(-\pf) \tilde{g}(\pf)- \tilde{g}(-\pf) \tilde{f}(\pf))$
vanishes for $p^+=0$ for all values of $\mathbf{p}_{\perp}$.  This
means that $[\phi (\tilde{f}), \phi (\tilde{g})]$ can be extended to
the light-front algebra, however the decomposition (\ref{f.8}),
and the algebraic normal ordering
discussed above is no longer well defined on the full light-front Fock
algebra.

In the case of the canonical field algebra, defining a vacuum
functional on the Weyl algebra uniquely determines the Hamiltonian
\cite{PhysRev.117.1137}\cite{Araki:1971yj} and hence the dynamics
needed to uniquely extend the algebra to the Local algebra.  This does
not happen in the light-front case.

In the light-front case the representations of the Weyl algebras
discussed above are unitarily equivalent \cite{Leutwyler:1970wn}  
for different mass fields.

\section{Equivalence}

In this section the meaning of equivalence of two field theories is
discussed.  We emphasize that it is important to distinguish the
equivalence of the theories and equivalence of the corresponding Weyl
representations.  The choice of field algebra plays an
important role for this
characterization.

The relevant algebra for a field theory is the local algebra, which is
distinguished from the light-front, canonical and Schlieder-Seiler
algebras by being closed under Poincar\'e transformations.  In
addition it contains operators that are localized in finite space-time
regions, which are needed to formulate locality conditions.  The GNS
construction using a Poincar\'e invariant vacuum functional leads to a
unitary representation of the Poincar\'e group on the GNS Hilbert
space.  Theories that have identical Wightman functions are unitarily
equivalent, since the Wightman functions are kernels of the Hilbert
space inner product which means that the correspondence between
the fields and vacuum in the Wightman functions preserves
all inner products.

It is possible for two theories to have Wightman functions that are
generally different, but nevertheless are identical on a sub-algebra.
An instructive example for the case of two-scalar fields with
different masses was given by Schlieder and Seiler
\cite{Schlieder:1972qr}.  
In this example they consider a sub-algebra of the local field algebra.
  
Let $\phi_1(x)$ and $\phi_2(x)$ be free scalar fields with
different masses, $m_1$ and $m_2$.  If $f(x)$ and $g(x)$ have Fourier
transforms $f(p)$ and $g(p)$ that agree on the mass shell for the
field of mass $m_1$, then $\phi_1 (f) = \phi_1 (g)$.  The condition
that two functions agree on a given mass shell divides the space of
test functions into disjoint equivalence classes of functions.

In general, if two test functions have Fourier transforms that agree
on one mass shell, their Fourier transforms are generally unrelated on
any other mass shell.  However out of the class of all test functions
there is a subspace of test functions, $f(p)$ that satisfy
\beq
{f (\sqrt{m_1^2 + \mathbf{p}^2},\mathbf{p}) \over   
(m_1^2 + \mathbf{p}^2)^{1/4}} =
{f (\sqrt{m_2^2 + \mathbf{p}^2},\mathbf{p}) \over   
(m_2^2 + \mathbf{p}^2)^{1/4}}.
\label{g.9}
\eeq
For test functions in this class, calculations imply that 
\beq
_1\langle 0 \vert \phi_1 (f) \phi_1 (g) \vert 0 \rangle_1 = 
_2\langle 0 \vert \phi_2 (f) \phi_2 (g) \vert 0 \rangle_2 .
\label{g.10}
\eeq
On this restricted sub-algebra the two-point functions of both fields
are identical.  In addition, the decomposition of this set of
functions into disjoint equivalence classes is identical for both
fields.  Finally, for free fields, every $n$-point function is a
product of two-point functions.  Because the two-point Wightman
functions for the different mass fields on this sub-algebra are
identical, the correspondence $\phi_1(f) \to \phi_2(f)$ and
$\vert 0 \rangle_1 \to \vert 0 \rangle_2$ is unitary.  On the other
hand this algebra is not invariant under Poincar\'e transformations.

While this is very restrictive class of test functions, it is still
large enough to be irreducible.  To see this note that
given any $g(p)$
(not necessarily in this class) there is a function $f_1(p)$ in this
class satisfying $\phi_1(f_1)= \phi_1(g)$ (they only have to have
Fourier transforms that agree on the mass shell).  There is also an
$f_2(p)$ in this class satisfying $\phi_2(f_2)= \phi_2(g)$.  However
there is no relation between $\phi_1(g)$ and $\phi_2(g)$ 
or $\phi_1(f_1)$ and $\phi_2(f_2)$.

Thus by limiting the space of test functions to a class that is not
closed under Poincar\'e transformations, one gets irreducible
unitarily equivalent representations of a sub-algebra of the full four
dimensional algebra for scalar fields with different mass.  
This equivalence is not preserved when the algebra is
extended to the local algebra.  

While the light-front algebra is not a sub-algebra of the
local algebra, it also has a limited set of test
functions that cannot distinguish fields of different mass, that are
however large enough to be irreducible.

Theories with different masses become inequivalent when the light-front 
Fock algebra is extended to the local algebra.  This will be discussed in 
the next section.

Another example that makes the role of the underlying algebra 
clear is comparing the two-point function of the local algebra
restricted to the light front to the two-point function constructed 
from the light-front Fock algebra.

The text book representation for the two-point function 
of a scalar field of mass $m$ in the 
local algebra can be found in \cite{bogoliubov:1959} 
\[
\langle 0 \vert \phi (x) \phi (y) \vert 0 \rangle =
\]
\[
-i {\epsilon (z^0) \delta (z_0^2 -\mathbf{z}^2) \over 4 \pi} 
+ {i m \theta (z_0^2 -\mathbf{z}^2)\over 
8 \pi \sqrt{(z_0^2 -\mathbf{z}^2)}}
(\epsilon (z^0)J_1 (m\sqrt{(z_0^2 -\mathbf{z}^2)})
- i N_1 (m\sqrt{(z_0^2 -\mathbf{z}^2)}))
\]
\beq
- { m \theta (\mathbf{z}^2- z_0^2)
\over 4 \pi^2 \sqrt{\mathbf{z}^2- z_0^2 }}
K_1 (m \sqrt{\mathbf{z}^2- z_0^2 })
\label{g.11}
\eeq
where $z^{\mu}= x^{\mu}-y^{\mu}$. Because this is Lorentz 
invariant, it is a function of $z^2$, so when $z^+=0$, there can 
be no $z^-$ dependence (except for the sign), and this becomes
\beq
\langle 0 \vert \phi (x) \phi (y) \vert 0 \rangle  \to 
-i {\epsilon (z^-) \delta (\mathbf{z}_{\perp}^2) \over 4 \pi} 
- { m 
\over 4 \pi^2 \sqrt{\mathbf{z}_{\perp}^2 }}
K_1 (m \sqrt{\mathbf{z}^2_{\perp} }).
\label{g.12}
\eeq
This can be compared to the direct construction 
of this quantity  
using the light-front Fock algebra, which knows nothing about the 
Lorentz symmetry.  The result is 
\beq
{1 \over (2 \pi)^3}
\int {d\tilde{\mathbf{p}}\theta (p^+) \over 2p^+} 
e^{i \tilde{\mathbf{p}}\cdot   \tilde{\mathbf{z}}}.
\label{g.13}
\eeq
This is a well-behaved distribution on the Schlieder-Seiler 
algebra, and it has a non-trivial dependence on $z^-$.  
On the other hand the Lorentz invariance of (\ref{g.12}) means that
it has no dependence on $z^-$.  Furthermore, (\ref{g.12}) is not 
even a distribution on the Schlieder-Seiler functions 
because the integral behaves like ${1 \over \mathbf{z}_{\perp}^2}$
at the origin \cite{Werner:2006}.  The difference in (\ref{g.12}) and (\ref{g.13})
is because the order of the light-front limit and integral matters.   

\section{Extension to the local algebra}

The explicit representation of the free scalar field in terms of the
light-front creation and annihilation operators, (\ref{f.1}), and the
light-front creation and annihilation operators in terms of the fields
restricted to the light front, (\ref{f.6}) and (\ref{f.7}), can be
combined to get the following  expression for the field on the local
algebra in terms of the field restricted to the light-front:
\beq
\phi (y)
= {1 \over 2 (2 \pi)^3} \int d\xf d\kf 
e^{-{i\over 2} {\mathbf{k}_{\perp}^2 + m^2 \over k^+}y^+ +i \kf\cdot
(\yf-\xf) }{\tilde{\phi}}(\xf) 
\label{h.1}
\eeq
where in this expression the $k^+$ integral is over {\it both positive
and negative values}.  This provides the desired extension from the 
light-front or Schlieder-Seiler algebras to the local algebra of the 
free field.

When $y^+=0$ this becomes a delta function in the light-front
coordinates and one recovers the field, ${\tilde{\phi}}(\xf)$, restricted to the
light front.  The specification of the 
mass $m$ in equation (\ref{h.1}) puts in the dynamical
information in $\phi (y)$.

Equation (\ref{h.1}) has the structure
\beq 
\phi (y) = \int d\xf F_m(y^+:\yf -\xf){\tilde{\phi}}(\xf). 
\label{h.2}
\eeq
If $f(y)$ is a Schwartz function in four-space-time variables 
and we define 
\beq
\gf_f (\xf) := \int d^4y  f(y)  F_m(y^+:\yf -\xf)
\label{h.3}
\eeq
then 
\beq
\phi (f) = \int \phi (x) f(x) d^4 x =
\int \gf_f (\xf) {\tilde{\phi}}(\xf) d\xf =
\tilde{\phi} (\gf_f) 
\label{h.4}
\eeq
where $\gf_f$ is a function of variables on the light-front hyper-plane.
This shows that the local free-field algebra can be expressed in terms
of the free-field light-front algebra.

The kernel $F_m(y:\yf -\xf)$ satisfies the Klein-Gordon equation for a
given mass.  Thus, it restores full Lorentz covariance.  The choice of $F_m$
also provides a dynamical distinction between free fields with
different masses.  It is responsible for the physical inequivalence of
free-field theories with different masses.  This inequivalence is
preserved if we use this to generate the canonical algebra by
integrating against test functions in three variables multiplied by
delta functions in time and their derivatives.

A free field theory is completely defined by its two-point Wightman
function.  
To compute the two-point Wightman function it is also necessary to
have the vacuum functional in addition to the algebra.

If the vacuum is annihilated by the light-front annihilation operator then
(\ref{h.2}) can be used to calculate the two-point Wightman 
function in terms of the restriction to the light front
\[ 
\langle 0 \vert \phi (x) \phi (y) \vert 0 \rangle =
\]
\[
{1 \over 8 (2 \pi)^9} \int d\xf_1 d\yf_1 d\kf d\pf 
{d\qf \over q^+} \theta (q^+)
e^{-i {\mathbf{p}_{\perp}^2 + m^2 \over 2 p^+}x^+ 
+i \pf \cdot  (\xf -\xf_1)}
e^{-i {\mathbf{k}_{\perp}^2 + m^2 \over 2 k^+}y^+ 
+i \kf \cdot (\yf- \yf_1) }
e^{i \qf \cdot (\xf_1 -\yf_1)} 
 =
\]
\[
{1 \over 2 (2 \pi)^3} \int {d\qf \over q^+} \theta (q^+) 
e^{-i {\mathbf{q}_{\perp}^2 + m^2 \over 2 q^+}(x^+-y^+) 
+i \qf (\xf - \yf)} =
\]
\beq
{1 \over (2 \pi)^3} \int \theta (q^+) \delta (q^2 + m^2) e^{i q \cdot (x-y)} d^4q
\label{h.5}
\eeq
which is the standard representation of the two-point Wightman function.

The interesting property of expression (\ref{h.5}) is that while the
$1/q^+$ denominator is log divergent for $q^+$ near zero, $e^{-i
  {\mathbf{q}_{\perp}^2 + m^2 \over q^+}x^+ }$ undergoes violent
oscillations in a neighborhood of $q^+=0$.  These oscillations
regularize the $1/q^+$ divergences.  To see that this happens note
that near the origin the integral has the same form as
\beq 
\int_0^a {e^{i c/q} \over q } dq =
\int_{c/a}^\infty {e^{iu }\over u } du = {\pi \over 2} -(Ci(c/a)+ i Si
(c/a)) 
\label{h.6}
\eeq 
where we have substituted $u=c/q$ and used equations
(5.5), (5.5.27) (5.5.26) in \cite{as} to get this result.  The
finiteness of this expression shows that the oscillations in the
exponent regulate the singularity at $q^+=0$.  It follows that 
the sharp restriction to the light front turns this 
``regulator'' off.  It is not something that can be continuously 
tuned on.

This shows that smearing with a test function in $x^+$
leads to an additional $p^+$ dependence that makes the
$1/p^+$ in the light-front inner product (\ref{n.19}) harmless.

To see this it is constructive to consider a test function 
that is a product of a test function in $x^+$ and a test function
in the light-front coordinates $\tilde{\mathbf{x}}$ of 
the form
\beq
f(x) = f_1(x^+)f_2 (\tilde{\mathbf{x}}).
\label{h.7}
\eeq
It follows that $\tilde{g}_f(\tilde{\mathbf{x}})$ defined by (\ref{h.3}) 
has a Fourier transform of the form
\beq
g_f(\tilde{\mathbf{p}}) = f_1({\mathbf{p}_{\perp}^2 + m^2 \over p^+})
f_2 (\tilde{\mathbf{p}}).
\label{h.8}
\eeq
Thus even if $f_2 (0,\mathbf{p}_{\perp})\not=0$,
if $f_1(p^-)$ is a Schwartz function, 
$g_f(\tilde{\mathbf{p}})$ will vanish faster than any 
power of $p^+$ as $p^+\to 0$.
  
This means that for free-field theories, $F_m$ maps all Schwartz test
functions in four variables into Schlieder-Seiler functions on the
light-front hyper-plane.  This uniquely fixes the light-front vacuum
by (\ref{f.14}).

To understand the significance of the mapping $f(x) \to
g_f(\tilde{\mathbf{x}})$ note that the operators
$\tilde{\phi}(\tilde{g}_f)$ for $f(x) \in {\cal S}(\mathbb{R}^4)$
generate a sub-algebra of the Schlieder-Seiler algebra.

On this sub-algebra we have the identity
\beq
_m\langle 0 \vert \phi(f_1) \cdots \phi (f_n) \vert 0 \rangle_m
= 
_{lf} \langle 0 \vert \tilde{\phi}(\tilde{g}_{f_1})
\cdots \tilde{\phi} (\tilde{g}_{f_n}) \vert 0 \rangle_{lf}
\label{h.9}
\eeq
which means that this correspondence preserves all Wightman
distributions on the local algebra.  This defines a unitary
mapping between the physical representation of the
local algebra and this representation of this sub-algebra of the
Schlieder-Seiler Fock algebra.

This unitary transformation maps the vacuum of the local
free field theory to the light front Fock vacuum.  If
$f_1$ and $f_2$ have space-like separated support then
this correspondence implies
\beq 
[\tilde{\phi} (\tilde{g}_{f_1}), \tilde{\phi} (\tilde{g}_{f_2})]=0 ,
\label{h.10}
\eeq
in addition $f_i(x) \to f_i'(x) = f_i(\Lambda x+a)$ implies that 
\beq
_{lf} \langle 0 \vert \tilde{\phi}(\tilde{g}_{f_1})
\cdots \tilde{\phi} (\tilde{g}_{f_n}) \vert 0 \rangle_{lf} =
_{lf} \langle 0 \vert \tilde{\phi}(\tilde{g}_{f'_1})
\cdots \tilde{\phi} (\tilde{g}_{f'_n}) \vert 0 \rangle_{lf} .
\label{h.11}
\eeq
These equations show how locality and a unitary representation of the
Poincar\'e group are realized in this light front-Fock representation
of this sub-algebra of the Schlieder-Seiler algebra.

Free fields of different mass involve different maps
that map to different sub-algebras of the Schlieder-Seiler algebra.

\section{dynamics}

In this section we discuss the extension of these results to the case
of interacting theories.  In particular we show how the local algebra
generated by the Heisenberg field operators of an interacting
theory can be mapped into a 
sub-algebra of the light-front Fock algebra.

The asymptotic completeness of the $S$ matrix means that the
theory has an irreducible set of asymptotic fields.  These are the
``in'' or ``out'' fields of the theory.  They are local free fields
with the masses of physical one-particle states of the theory.  In
general there may also be local asymptotic fields \cite{Nishijima}
for composite particles.

In \cite{Haag:1955ev}\cite{Glaser:1957} it is shown under mild
assumptions that any linear operator $A$ on the Hilbert space can be
expanded as a series of normal products of asymptotic fields.  This
expansion is referred to as the Haag expansion \cite{Greenberg:1965}.
For the Heisenberg field of an interacting theory the Haag expansion 
has the form:
\beq
\phi (x) = \sum_{n=0}^\infty {1 \over n!}
\int d^4x_1 \cdots d^4x_n L_n (x; x_1, \cdots ,x_n) 
:\phi_{in} (x_1) \cdots \phi_{in} (x_m).
\label{dyn.1}
\eeq
This expansion is non-trivial - the masses in asymptotic fields
are physical particle masses and in general there are asymptotic 
fields for both composite and elementary particles.   

The Poincar\'e covariance of the Heisenberg and asymptotic fields 
means that the coefficient functions
$L_n (x; x_1, \cdots ,x_n)$ are invariant 
\beq
L_n (x; x_1, \cdots ,x_n) =
L_n (\Lambda x +a ; \Lambda x_1+ a , \cdots ,\Lambda x_n+a)
\label{dyn.2}
\eeq
under Poincar\'e transformations $(\Lambda,a)$.  For higher spin
composite fields the invariance is replaced by an obvious
covariance.

Furthermore,  if the Heisenberg field $\phi (x)$ is an operator-valued
tempered distribution, 
then $\phi (f)$ should be a Hilbert space operator when 
$f(x)$ is Schwartz function.  For the Hagg expansion of $\phi (f)$
to also be an operator,  the smeared coefficient functions
\beq
L_n(f,x_1, \cdots ,x_n) :=
\int f(x) L_n (x; x_1, \cdots ,x_n) d^4 x  
\label{dyn.3}
\eeq
should behave like Schwartz test functions in $4n$ variables,
since the asymptotic fields are all operator valued tempered
distributions.

Using, (\ref{h.1}-\ref{h.2}), each of the asymptotic fields can be
expressed as the extension of a light-front field, using the kernels
$F_m (x^+;\tilde{\mathbf{x}}-\tilde{\mathbf{y}})$.  Using these in the
Haag expansion gives the following representation of the Heisenberg
field  in terms of products of algebraically normal-ordered fields
restricted to a light front:
\[
\phi (x) = \sum_{n=0}^\infty {1 \over n!}
\int d^4x_1 \cdots d^4x_n L_n (x; x_1, \cdots ,x_n)
F_{m_1}(x_1^+;\tilde{\mathbf{x}}_1-\tilde{\mathbf{y}}_1) 
\cdots 
F_{m_n}(x_n^+;\tilde{\mathbf{x}}_n-\tilde{\mathbf{y}}_n)
\times
\]
\beq
d\tilde{\mathbf{y}}_1 \cdots d\tilde{\mathbf{y}}_n
:\tilde{\phi}_{10} (\tilde{\mathbf{y}}_1)
\cdots
\tilde{\phi}_{n0} (\tilde{\mathbf{y}}_n):
\label{dyn.4}
\eeq

If $\phi (x)$ is smeared with a Schwartz function, $f(x)$,  
and $L_n(f,x_1, \cdots ,x_n)$ is a Schwartz function in 
$4n$ variables, then the $p^+=0$ behavior of the light-front
fields will be suppressed in (\ref{dyn.4})
by the mechanism (\ref{h.6}).
In this representation the vacuum functional is fixed by (\ref{f.14}),
as it is in the case of the free field.  The non-trivial aspects of the 
dynamics appear in the extension for the full Heisenberg algebra.

It follows that elements of the local Heisenberg algebra can be
expressed as elements of the light-front Fock algebra
\beq
\phi (f) = \sum_{1 \over n!}
\int \tilde{L}_n (f,
\tilde{\mathbf{y}}_1, \cdots , \tilde{\mathbf{y}}_n)
d\tilde{\mathbf{y}}_1 \cdots d\tilde{\mathbf{y}}_n
:\tilde{\phi}_{10} (\tilde{\mathbf{y}}_1)
\cdots
\tilde{\phi}_{n0} (\tilde{\mathbf{y}}_n):
\label{dyn.5}
\eeq
where 
\beq
\int \tilde{L}_n (f,
\tilde{\mathbf{y}}_1, \cdots , \tilde{\mathbf{y}}_n) = 
\int d^4x_1 \cdots d^4x_n L_n (f; x_1, \cdots ,x_n)
F_{m_1}(x_1^+;\tilde{\mathbf{x}}_1-\tilde{\mathbf{y}}_1) 
\cdots 
F_{m_n}(x_n^+;\tilde{\mathbf{x}}_n-\tilde{\mathbf{y}}_n).
\label{dyn.6}
\eeq
As in the free field case, this correspondence generates a unitary
mapping from the local algebra of the Heisenberg field to a
sub-algebra of the Schlieder-Seiler algebra.  

Specifically this correspondence has the form
\beq
\phi (f) \to \tilde{A}[f]
\eeq
where $\tilde{A}[f]$ is the right hand side of (\ref{dyn.5}), 
which is an element of the light-front Schlieder
Seiler algebra.  The correspondence
\beq 
\langle 0 \vert \phi(f)_1 \cdots \phi (f_n) \vert 0 \rangle = 
_{lf}\langle 0 \vert A[f_1] \cdots A[f_n] \vert 0 \rangle_{lf}  
\eeq
defines a unitary map from the Hilbert space generated
by the local field $\phi (f)$ to the Hilbert space generated 
by the $A[f]$ on the light-front Fock vacuum.

This correspondence relates the Heisenberg vacuum to the 
light-front Fock vacuum.  It
preserves local commutation relations,  the sub-algebra is
Poincar\'e invariant, and the Fock vacuum is also Poincar\'e invariant
on this sub-algebra.

In this case all of the dynamical information is contained in the
mapping which defines the sub-algebra of the light front
Fock algebra.  All of the dynamical information is removed from the
vacuum. 

This discussion does not directly apply to QCD.  In the case of QCD the
asymptotic fields are composite and color singlets.  See ref
\cite{Greenberg:1978} for a discussion.
While these
asymptotic fields generate the physical Hilbert space, they do not
generate the non-singlet part of the Hilbert space.  One possible
advantage of the expansion (\ref{dyn.2}) is that the light-front
fields that appear in the expansion do not carry information about
asymptotic masses, so they might provide a means to extend the light
front ``Haag expansion'' (\ref{dyn.2}) to the non-singlet sectors, where there
are no asymptotic fields.

\section{zero modes}

In the previous section we demonstrated that it was possible to express a
smeared interacting Heisenberg field as an expansion in terms of
algebraically normal ordered free fields restricted to a light front,
where the vacuum is the free light-front Fock vacuum.  In this
expansion contributions associated with $p^+=0$ are suppressed by the
coefficient functions, $\tilde{g}_n (f,\tilde{\mathbf{x}}_1 \cdots
\tilde{\mathbf{x}}_n)$ which are Schlieder-Seiler functions on the
light front.  

Unfortunately the operators that appear in dynamical equations, like
the Poincar\'e generators,  involve local products of fields rather
than products of smeared fields.  A characteristic property of any
quantum field theory is that local products of local fields are not
defined.  This is because the fields are operator-valued
distributions. 

The leading term in the Haag expansion is  
\beq
\phi (x) = Z \phi_{in}(x) + \cdots =
Z \int
F_{m}(x^+;\tilde{\mathbf{x}}-\tilde{\mathbf{y}}
\phi_0(\tilde{\mathbf{y}})
d\tilde{y} + \cdots 
\label{zm.1}
\eeq
where $Z$ is a constant that relates the normalization of $\phi (x)$
to the normalization of the asymptotic field.  

This means that singularities of local products of Heisenberg fields
on the light front are determined in part by the singularities of the 
corresponding local products of asymptotic fields on the light front:
\beq
\phi (x)^n \to Z^n \phi_{in}(x)^n + \cdots
\eeq
What is relevant is that local products of extensions of the 
fields do not suppress $p^+=0$ singularities.

Recall that previously we showed that for Schlieder-Seiler test functions
the vacuum was given by a functional of the form
\beq
E[f] := 
\langle 0 \vert e^{i \tilde{\phi}(\ff)} \vert 0 \rangle =
\langle 0 \vert :e^{i \phi(\ff)}: \vert 0 \rangle
e^{{1 \over 2} (\ff,\ff)}.
\eeq
If we expand this out 
\beq
\langle 0 \vert e^{i \tilde{\phi}(\ff)} \vert 0 \rangle =
\sum_{m=0}^\infty {i^n \over n!} \langle 
0 \vert \tilde{\phi} (\ff)^n \vert 0 \rangle 
\eeq
all of the fields are smeared with Schlieder-Seiler test functions
before computing the vacuum expectation value.  

Removing the test functions by taking functional derivatives with
respect to the Schlieder-Seiler functions gives kernels that represent
Schlieder-Seiler distributions.  These are not sensitive to what
happens at $p_i^+=0$ for Schlieder-Seiler functions.  In order to deal
with local operator products the vacuum functional needs to be
extended to treat points with $p^+=0$.  The two-point function on the
light front in the light front algebra is up to a constant the
ill-defined light front scalar product.

This has a logarithmic singularity at $p^+=0$.  This scalar product
can be renormalized in a number of ways.  For example 
\beq 
(\ff,\ff) \to
\int
{dp^+ d\mathbf{p}_{\perp} \theta (p^+)
  \over p^+ } \ff(-\tilde{p})\ff(\tilde{p})
- 
{dp^+ d\mathbf{p}_{\perp} \theta (p^+)
\over p^+ } \ff(0,-\mathbf{p}_{\perp})\ff(0,\mathbf{p}_{\perp})
e^{-\beta p^+}
\eeq
This recovers the expected result on Schlieder-Seiler functions, but it is
well-defined on test functions that do not vanish at $p^+=0$.  One
apparent problem with this renormalization is that it breaks boost
invariance in the $\hat{\mathbf{n}}$ direction.  This is not critical
because the mapping $\tilde{g}_n(x,\tilde{y}_1, \cdots, \tilde{y}_n )$ 
does not preserve 
kinematic covariance when $x^+\not=0$.  The full invariance must be 
recovered in the final extension to the local algebra.

Once test functions that do not vanish at $p^+=0$ are included,
it is possible to have
\beq
\langle 0 \vert :e^{i\tilde{\phi}(\tilde{f})}: \vert 0 \rangle
\not=1 .
\eeq
While the algebraic normal ordering means that this is
$1$ for functions with positive $p^+$ support, One can extend
this functional to include additional contributions
that are concentrated at $p^+=0$.  Some of the allowed
contributions have the form 
\[
\langle 0 \vert :e^{i\tilde{\phi}(\tilde{f})}: \vert 0 \rangle
=
\sum {i^n \over n!} 
\langle 0 \vert :\tilde{\phi}(\tilde{f})^n: \vert 0 \rangle
=
e^{\sum {i^n \over n!}  
\langle 0 \vert :\tilde{\phi}(\tilde{f})^n: \vert 0 \rangle_c}
\]
\beq
=
e^{\sum {i^n \over n!}  
\int w_n^c (\mathbf{p}_{1\perp}, \cdots ,\mathbf{p}_{n\perp}) 
\ff (0, \mathbf{p}_{1\perp}) \cdots \ff (0, \mathbf{p}_{n\perp})
\delta (\sum \mathbf{p}_{n\perp}) 
d\mathbf{p}_{1\perp} \cdots d\mathbf{p}_{n\perp}
}
\label{gns.14xx}
\eeq
where the subscript $c$ indicates the connected part of the $n$-point
function which is defined in terms of the distributions of the form
\beq
w_n^c (\mathbf{p}_{1\perp}, \cdots ,\mathbf{p}_{n\perp})
\eeq
This  does not exhaust the possible zero mode
contributions;  distributions of the form 
\beq
w_n^c(\ff, \cdots \ff) =
\int w_n (\mathbf{p}_{1\perp}, \cdots \mathbf{p}_{n\perp},\xi_1,\cdots \xi_n )
\delta (\sum_i p_i^+)
\delta (\sum \xi_j -1) \prod \theta (\xi_k)
\prod_{i=1}^n  f(\pf_i)  d\pf_i .
\label{gns.18}
\eeq
where the $\xi_k$ light front momentum fractions,
can also be 
considered.  What is relevant is that the ``measure'' is kinematically 
invariant.  

It is clear the that need for zero modes is related to renormalization
of operator products.  This is non trivial because both ultraviolet
and infrared singularities must be treated together in a manner that
preserves the rotational and $\hat{n}$-boost covariance of the theory
and preserves the positivity of the Hilbert space inner product.  The
presence of zero modes defines an extension of the light-front Fock
vacuum.  These extensions may be needed in perturbative expansions of
expressions that involve local products of fields.  While they may
play a role in constructing the coefficient functions in the
light-front Haag expansion, they do not directly contribute to the
final representation of the smeared Heisenberg operators.

\section{Summary}

The goal in this paper was to understand (1) why the light-front
vacuum of an interacting theory is the same as the Fock vacuum while 
the vacuua differ in the conventional formulation of field theory and 
(2) the role of zero 
modes in the light-front vacuum.  

The first issue involves determining the meaning of the vacuum.  For
free fields the characterization of the vacuum by an annihilation
operator is incomplete.  The physically relevant characterization of
the vacuum is as a positive linear functional of an algebra of field
operators.

When the vacuum is characterized as a linear functional on an algebra
of operators, the choice of algebra matters.  In order to realize
Poincar\'e symmetry manifestly, or localize field observables to finite
regions of space time, the space of test functions of the algebra should 
include functions with support in finite space-time volumes and
should be invariant under space-time translations and Lorentz
transformations. 

From a physics point of view the relevant algebra is generated by
fields smeared with test functions of four space-time variables.  
This algebra is Poincar\'e invariant.  

In this work we showed that the local algebra of both free and
interacting fields could be mapped into sub-algebras of the
Schlieder-Seiler algebra.  We also showed that the vacuum is trivial
and uniquely defined on the Schlieder-Seiler algebra.  This mapping
moves the dynamics in the light-front vacuum into the mapping.  The
mapping defines a unitary transformation from the physical
representation of local algebra generated by a scalar field to a
sub-algebra of the Schlieder-Seiler algebra with the Fock vacuum.
This unitary correspondence leads to a formulation of locality and
Poincar\'e invariance on a subspace of the light front Fock space.
Sub-algebras of the Schlieder-Seiler algebra associated with different
local fields are not unitarily related.

The Schlieder-Seiler algebra has no place for zero modes, but
extension to include zero modes may be required to treat local
operator products that arise in perturbation theory.  A large
class of zero-mode contributions are possible, but they are restricted 
by positivity and Poincar\'e invariance.

\begin{acknowledgments}
This work was performed under the auspices of the U. S. Department of Energy,
Office of Nuclear Physics, under contract No. DE-FG02-86ER40286 with the
University of Iowa.
\end{acknowledgments}

\end{document}